\documentclass[journal,onecolumn]{IEEEtran}

\usepackage{amsmath,amssymb,amscd,latexsym,dsfont}
\usepackage{booktabs}
\usepackage{cite}

\usepackage{multicol}
\usepackage{array}
\usepackage{wrapfig}

\usepackage{tikz}  						
\usetikzlibrary{shapes.geometric, arrows,calc}
\usetikzlibrary{positioning}
\usetikzlibrary{decorations.pathreplacing}
\tikzstyle{arrow} =[thick,->,>=stealth] 

\usepackage{pgfplots}
\pgfplotsset{compat=newest}
\usetikzlibrary{plotmarks}
\usetikzlibrary{arrows.meta}
\usepgfplotslibrary{patchplots}
\usepackage{grffile}
\usepackage{graphicx}

\newcommand\blfootnote[1]{%
  \begingroup
  \renewcommand\thefootnote{}\footnote{#1}%
  \addtocounter{footnote}{-1}%
  \endgroup
}

\begin{document}
\clearpage

\title{A Data-driven Optimization of First-order Regular Perturbation Coefficients for Fiber Nonlinearities}

\author{Astrid~Barreiro,~\IEEEmembership{Student Member,~IEEE,}
        Gabriele~Liga,~\IEEEmembership{Member,~IEEE,}
        and~Alex~Alvarado,~\IEEEmembership{Senior Member,~IEEE}
}

\maketitle

\begin{abstract}
We study the performance of gradient-descent optimization to estimate the coefficients of the discrete-time first-order regular perturbation (FRP). With respect to numerically computed coefficients, the optimized coefficients yield a model that (i) extends the FRP range of validity, and (ii) reduces the model’s complexity.
\end{abstract}

\IEEEpeerreviewmaketitle

\section{Introduction}
The Manakov equation (ME) describes the propagation of signals in an optical fiber. The framework of perturbation theory is widely employed in fiber-optic transmission systems to provide compact analytical approximations to the ME solution \cite{Vannucci2002b}. The most popular approximation is based on the regular perturbation expansion on the nonlinear parameter $\gamma$ truncated to the first-order term. We refer to this method as first-order regular perturbation (FRP). 

FRP in discrete-time was extensively studied in \cite{Mecozzi2012,DarJLT2015}. Two main pitfalls can be highlighted in discrete-time FRP models: (i) decreased accuracy as transmitted power increases, and (ii) high complexity in the evaluation of the model. The first pitfall is due to the fact that nonlinearities cannot be considered a perturbation at high powers. The second pitfall is due to the generally large number of perturbation coefficients (also referred to as \emph{kernels}) required to achieve good accuracy. This number grows exponentially with the effective memory of the channel, which in turn increases when bandwidth or fiber length increase. The two aforementioned drawbacks therefore restrict the suitability of FRP to linear or pseudo-linear transmission scenarios, as well as to small bandwidth and short distances.

Improved channel models have been proposed to enhance the accuracy of FRP-based models at high powers \cite{Vannucci2002b,Oliari2020c,Secondini2015b}. Among others, the FRP computational complexity has been reduced by using the temporal phase-matching symmetry \cite{Tao}, or by imposing a quantization of the coefficients \cite{Peng2015}. Machine learning algorithms have also been proposed to avoid computing the FRP coefficients \cite{Zhang2019}. However, the approaches above have not been used to extend the FRP validity region.

In this paper, we propose a numerical approach to tackle the FRP drawbacks via a data-driven optimization of the FRP coefficients. The optimization is based on a complex-variable gradient descent algorithm. By using the optimized coefficients, our method provides an extension of the validity region of the FRP when using a given model memory size, or alternatively, it relaxes the model memory requirements for a target accuracy at fixed transmitted power. For the considered system, FRP shows good accuracy up to 10~dBm, while our approach offers the same accuracy 6~dB beyond that.

\section{Theory and Simulation Results}
\begin{wrapfigure}{r}{0.31\textwidth}
  \centering 
    \resizebox{5.5cm}{!}{\tikzstyle{simple_box} = [rectangle, rounded corners, minimum width=0.1cm, minimum height=1cm, text centered, text width=1.5cm, draw=black]
\tikzstyle{simple_box_b} = [rectangle, rounded corners, minimum width=0.5cm, minimum height=.7cm, text centered, text width=1cm, draw=black]
\tikzstyle{simple_box_c} = [rectangle, rounded corners, minimum width=0.1cm, minimum height=0.7cm, text centered, text width=5cm, draw=black]

\begin{tikzpicture}
\draw[dashed, fill=orange!10] (-4.43,-2.1) rectangle ++ (6.1,3.15);
\node (label) at (-1.3,1.27) {FRP approximation in \eqref{eq:single_ch_dual_pol_DRP}};

\node (ssmf) [simple_box,fill=white,text width=1.25cm] {Optical\\Channel};
\node[left=1cm of ssmf] (bo1) {\large{$\bigotimes$}};
\node (modformat) [simple_box,left=2cm of ssmf,text width=1.25cm,fill=white] {MOD};
\node (cdc_b) [simple_box, below=0.5 of ssmf,text width=1.0cm,xshift=0.5cm,fill=white] {CDC};
\node (mf_b) [simple_box, left=0.5 of cdc_b,text width=1.0cm,fill=white] {MF};
\node[below=0.7cm of modformat, xshift=-.1cm] (bo2) {\large{$\bigotimes$}};

\node (samp_l_b) [coordinate, left=0.4cm of mf_b] {};
\node (samp_r_b) [coordinate, left=.6cm of samp_l_b] {};
\draw [arrow] (mf_b) -- (samp_l_b);
\draw [thick] (samp_l_b) -- (-2.7,-1.08);
\draw [thick,->] (-2.3,-1.1) arc (100:190:13pt);
\node (T) at (-2.6,-1.8) {\scriptsize{$0,T,2T, \ldots$}};
\draw [arrow] (samp_r_b) -- (-3.4,-1.515);

\node[left=0.4cm of modformat] (insymb) {$\underline{\boldsymbol{a}}$};
\node [below=1.04cm of insymb] (yout) {$\underline{\boldsymbol{r}}$};
\draw [arrow] (insymb) -- (modformat);
\node (n1) at (-2.2,0) {};
\draw [arrow] (modformat)--(n1);
\node (n2) at (-2,0) {};
\draw [arrow] (n2)-- node[anchor=south] {$\boldsymbol{A}(t,0)$} (ssmf);
\coordinate (sco1) at (1.5,0);
\draw [thick] (ssmf) -- (sco1);
\draw [arrow] (sco1)|- node[anchor=east,yshift=0.72cm] {$\boldsymbol{A}(t,L)$} (cdc_b);
\draw [arrow] (cdc_b)-- (mf_b);
\draw [arrow] (-3.82,-1.515) -- (yout);

\node [above=.15cm of bo1] (g1) {$\sqrt{E_s}$};
\coordinate (supbo1) at (-2.11,0.22);
\draw [arrow] (g1) -- (supbo1);
\node [above=.15cm of bo2] (g2) {$1/\sqrt{E_s}$};
\coordinate (supbo2) at (-3.6,-1.32);
\draw [arrow] (g2) -- (supbo2);

\end{tikzpicture}}
  \caption{{\footnotesize System model under study.}} 
  \label{fig:system_model}
\end{wrapfigure}
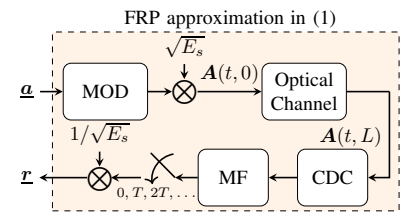

In this work, we consider a single-channel, single-span fiber-optic transmission system using a polarization-multiplexed signal, as shown in Fig.~\ref{fig:system_model}. A sequence of two-dimensional complex symbols with unit energy per complex dimension $\underline{\boldsymbol{a}}=\ldots, \boldsymbol{a}_{n-1},\boldsymbol{a}_{n},\boldsymbol{a}_{n+1},\ldots$ linearly modulates a pulse shape $h(t)$ to generate the power-scaled transmitted signal $\boldsymbol{A}(t,0)$, which is propagated through a length-$L$ (baseband equivalent) optical channel modeled by the attenuation-normalized ME. At the receiver side, ideal chromatic dispersion compensation is performed, followed by matched-filter and sampling at the symbol rate, yielding a sequence of received 2D complex vectors $\underline{\boldsymbol{r}}=\ldots, \boldsymbol{r}_{n-1},\boldsymbol{r}_{n},\boldsymbol{r}_{n+1},\ldots$ \blfootnote{\emph{Notation Convention:} Underline bold symbols denote sequences of symbols, sets are denoted by calligraphic letters, and $()^\dagger$ denotes Hermitian conjugate.}\blfootnote{\emph{Acknowledgments:} The work of A. Barreiro and A. Alvarado has received funding from the European Research Council (ERC) under the European Union's Horizon 2020 research and innovation programme (grant agreement No 757791). The work of G. Liga has received funding from the EuroTech Postdoc programme under the European Union’s Horizon 2020 research and innovation programme (Marie Skłodowska-Curie grant agreement No. 754462).}

For small enough values of the nonlinear coefficient $\gamma$, FRP approximates the exact solution to the ME yielding the following input-output relation in discrete-time:
\begin{align}\label{eq:single_ch_dual_pol_DRP}
	\boldsymbol{r}_n\approx \boldsymbol{a}_n + \jmath \frac{8}{9} \gamma E_S\sum_{(k,l,m)\in\mathcal{S}} \boldsymbol{a}_{n+k}^\dagger \boldsymbol{a}_{n+l}\boldsymbol{a}_{n+m}~S_{klm},
\end{align}
where $E_s$ is the average energy per transmitted symbol, $\jmath$ the imaginary unit, and $S_{klm}$ are the complex coefficients modeling self-phase modulation, defined as \cite[eq. (4)]{DarJLT2015}
 \begin{align}
	S_{klm} = &\int_0^L e^{-\alpha z} \int_{-\infty}^{\infty} \hspace{0.1cm} h^*(z,t) h^*(z,t-kT)h(z,t-lT)h(z,t-mT)\,\mathrm{d}t\mathrm{d}z,
	\label{eq:SMP_kernel}
 \end{align}
 where $h(z,t)$ is the dispersed pulse $h(t)$ at distance $z$, and $T$ is the symbol duration.
Since the $S_{klm}$ coefficients have an exponentially decaying magnitude with respect to $S_{000}$, the original infinite sums in the FRP model can be truncated to a subset of $\mathbb{Z}^3$:  $\mathcal{S}\triangleq \{(k,l,m)\in\mathbb{Z}^3:-M\leq k,l,m \leq M\}$, where $M$ denotes the model's memory. In \eqref{eq:single_ch_dual_pol_DRP}, $(2M+1)^3$ $S_{klm}$ coefficients must be computed. 

\begin{figure*}[!t]
 \centering 
 \resizebox{0.34\textwidth}{!}{\tikzstyle{simple_box} = [rectangle, rounded corners, minimum width=0.1cm, minimum height=1cm, text centered, text width=1.5cm, draw=black]

\begin{tikzpicture}
\draw[dashed, fill=orange!12,rounded corners] (-5.3,0) rectangle ++ (9.6,1.5);
\draw[dashed, fill=green!12,rounded corners] (-5.3,-2.25) rectangle ++ (9.6,1.4);
\draw[dashed, fill=purple!12,rounded corners] (-5.3,-4.2) rectangle ++ (9.6,1.5);

\node (ssmf) [simple_box,fill=white,text width=2.2cm,yshift=.75cm] {System model \\Fig. \ref{fig:system_model}};
\node (insymb) [left=2cm of ssmf] {$\underline{\boldsymbol{a}}$};
\node (outsymb) [right=2.4cm of ssmf] {$\underline{\boldsymbol{r}}$};

\node (nbgd) [simple_box,fill=white,below=1.35cm of ssmf,text width=1.5cm,xshift=-2cm] {NBGD\\Optimizer};
\node (frp) [simple_box,fill=white, right=2cm of nbgd,text width=1.25cm] {FRP \\ in \eqref{eq:single_ch_dual_pol_DRP}};
\node (ins) [left=0.5cm of nbgd] {\small{$\tilde{S}_{klm}^{(l)}$}};
\node (estsymb) [right=0.3cm of frp] {$\tilde{\underline{\boldsymbol{r}}}^{(l+1)}$};

\node (nbgd_in) [above=.2cm of nbgd] {$\underbrace{\{\underline{\boldsymbol{a}},\underline{\boldsymbol{r}}\}}$};
\node (frp_in) [above=.2cm of frp] {$\underbrace{\underline{\boldsymbol{a}}}$};

\node (rmse) [simple_box,fill=white, below=.9cm of frp,text width=1.0cm] {RMSE};
\node (thre) [simple_box,fill=white, left=1.5cm of rmse,text width=1.5cm] {Threshold\\check};
\node (end) [left=2cm of thre, fill=white,text width=.5cm,rectangle, rounded corners, minimum width=.8cm, minimum height=.5cm, text centered, draw=black] {End};

\draw [arrow] (insymb) -- (ssmf);
\draw [arrow] (ssmf) -- (outsymb);
\draw [arrow] (ins) -- (nbgd);
\draw [arrow] (nbgd) -- node[anchor=south] {$\tilde{S}_{klm}^{(l+1)}$} (frp);
\draw [arrow] (frp) -- (estsymb);
\draw [arrow] (rmse) -- (thre);
\node(coo0) [coordinate, below=1.5cm of ins] {};
\node(coo01) [left=1.05cm of thre] {$\diamond$};
\draw [arrow,dashed] (coo0) -- node[anchor=west,yshift=.2cm] {Increment $l$} (ins);
\draw [arrow] (thre)  -- (coo01);
\draw [arrow] (nbgd_in) -- (nbgd);
\draw [arrow] (frp_in)  -- (frp);
\draw [arrow,dashed] (coo01)  -- node[anchor=north] {\small{yes}} (end);
\node (no) [left=1cm of thre,yshift=.4cm] {\small{no}};

\draw [arrow] (estsymb) |- (rmse.20);
\draw [arrow] (outsymb) |- (rmse.340);
\node(coo1) [coordinate, above=1.3cm of ins] {};
\node(coo2) [coordinate, above=1.3cm of estsymb] {};
\draw [arrow] (coo1) |- (nbgd_in);
\draw [arrow] (coo2) |- (frp_in);

\node (lab1) at (-4.9,1.2) {(i)};
\node (lab2) [below=1.8cm of lab1] {(ii)};
\node (lab3) [below=1.25cm of lab2] {(iii)};

\node[font=\Large] (lab1) at (-5.8,0.7) {(a)};
\node[font=\Large] (cap0) at (0,-5.3) {NBGD block diagram};
\node (cap1) at (0,-5.6) {};  

\end{tikzpicture}} 
 \hfill
 \resizebox{0.32\textwidth}{!}{\definecolor{mycolor1}{rgb}{0.07451,0.62353,1.00000}%
\definecolor{mycolor2}{rgb}{0.71765,0.27451,1.00000}%
\definecolor{mycolor3}{rgb}{0.39216,0.83137,0.07451}%
\begin{tikzpicture}
\begin{axis}[%
every axis/.append style={font=\normalsize},
width=0.4\columnwidth,
height=0.32\columnwidth,
xmin=0,
xmax=20,
xticklabel style={font=\normalsize},
xtick={0,5,10,15,20},
xlabel style={font=\color{white!15!black}},
xlabel={Input power~[dBm]},
ymin=0,
ymax=45,
ytick={ 0, 10, 20, 30, 40, 50},
ylabel style={font=\color{white!15!black}},
yticklabel style={font=\normalsize},
ylabel={Nonlinear SNR~[dB]},
axis background/.style={fill=white},
grid style={dashed,lightgray!75},
xmajorgrids,
ymajorgrids,
legend style={at={(0.72,0.785)}, anchor=center, legend cell align=left, align=left, legend columns=1,font=\footnotesize}
]
\draw[fill=red!1,line width=0.1mm] (14.5,8) rectangle (17.5,17.5);

\addplot [smooth, color=red, line width=1pt, mark=*, mark options={solid, red,fill=white}]
  table[row sep=crcr]{%
-5	53.6857573803566\\
-4	51.6497797588019\\
-3	49.688721745459\\
-2	47.536875804605\\
-1	45.6337055315739\\
0	43.6321543559351\\
1	41.5754209221533\\
2	39.6764603573661\\
3	37.5835221680073\\
4	35.6295086940571\\
5	33.5890789311051\\
6	31.6109563467305\\
7	29.5621150597579\\
8	27.5641350790601\\
9	25.5288854263801\\
10	23.4646565731038\\
11	21.4626742964967\\
12	19.4242299406946\\
13	17.3592236397748\\
14	15.3405224497856\\
15	13.1487682586546\\
16	10.9854620881499\\
17	8.70042157033346\\
18	6.42305110234847\\
19	3.85599772270999\\
20	1.05642683294964\\
};
\addlegendentry{SSFM}

\addplot [color=mycolor3, line width=1pt, mark size=2pt, mark=square*, mark options={solid, mycolor3,fill=white}]
  table[row sep=crcr]{%
0	43.4787056901927\\
1	41.4357126573027\\
2	39.4511585485523\\
3	37.4451331018825\\
4	35.4967902321854\\
5	33.4637534515485\\
6	31.5187044958935\\
7	29.5099748952219\\
8	27.5744871673083\\
9	25.6289646762613\\
10	23.7073851239591\\
11	21.8544611096726\\
12	20.1312017165465\\
13	18.4475247566181\\
14	16.8768121214882\\
15	15.5822833595708\\
16	14.4771752542635\\
17	13.6044568665678\\
18	12.9211292114654\\
19	12.4734822698994\\
20	12.1163717162127\\
};
\addlegendentry{FRP$_{\text{INT}}$ $M = 5$}

\addplot [smooth, color=mycolor1, line width=1pt, mark size=2pt, mark=square*, mark options={solid, mycolor1,fill=white}]
  table[row sep=crcr]{%
-5	53.5457756590428\\
-4	51.549497753753\\
-3	49.5716161201667\\
-2	47.5647919256851\\
-1	45.5820208221136\\
0	43.5408640890414\\
1	41.5681633799158\\
2	39.5582785622158\\
3	37.5675785690716\\
4	35.5525509812604\\
5	33.6165199480068\\
6	31.62642010869\\
7	29.6599792346274\\
8	27.7074850686542\\
9	25.8022014615331\\
10	23.9704141160926\\
11	22.155628553276\\
12	20.5366068421904\\
13	18.9895313949176\\
14	17.6829176277617\\
15	16.5602375573703\\
16	15.6781381321746\\
17	14.974631724909\\
18	14.5112308750031\\
19	14.1855610531149\\
20	13.9100366356101\\
};
\addlegendentry{FRP$_{\text{INT}}$ $M = 15$}

\addplot [smooth,color=mycolor2, line width=1pt, mark=otimes*, mark options={solid, mycolor2,fill=white,scale=1.2}]
  table[row sep=crcr]{%
0	43.606589928599\\
1	41.6011068045892\\
2	39.5767265999009\\
3	37.6254791568369\\
4	35.5807911756898\\
5	33.6085335866393\\
6	31.57625094641\\
7	29.5735912949339\\
8	27.6045669603944\\
9	25.5128818188214\\
10	23.4948679508265\\
11	21.4640857542099\\
12	19.4816079924955\\
13	17.4104851955014\\
14	15.3102576560053\\
15	13.2939228510438\\
16	11.2025109010669\\
17	9.07528878393829\\
18	7.01571008104705\\
19	4.80449812631065\\
20	2.79013332363904\\
};
\addlegendentry{FRP$_{\text{NBGD}}$ $M = 5$}


\node[font=\normalsize] at (1.2,37) {(b)};
\coordinate (schema) at (axis cs:5,13);
\draw[line width=0.1mm] (axis cs:9.2,23.2) -- (axis cs:17.5,17.5);
\draw[line width=0.1mm] (axis cs:9.2,5) -- (axis cs:17.5,8);

\draw[dashed,thick] (axis cs:0,0)--(axis cs:0,47); 
\draw[dashed,thick] (axis cs:10,0)--(axis cs:10,50);
\draw[dashed,thick] (axis cs:16,0)--(axis cs:16,50); 

\draw[<->,thick] (axis cs:10,3)--(axis cs:16,3); 
\node[font=\small] at (13.5,4.7) {\footnotesize $6$~dB};

\end{axis}

\node[] at (schema){\resizebox{2.7cm}{!}{\input{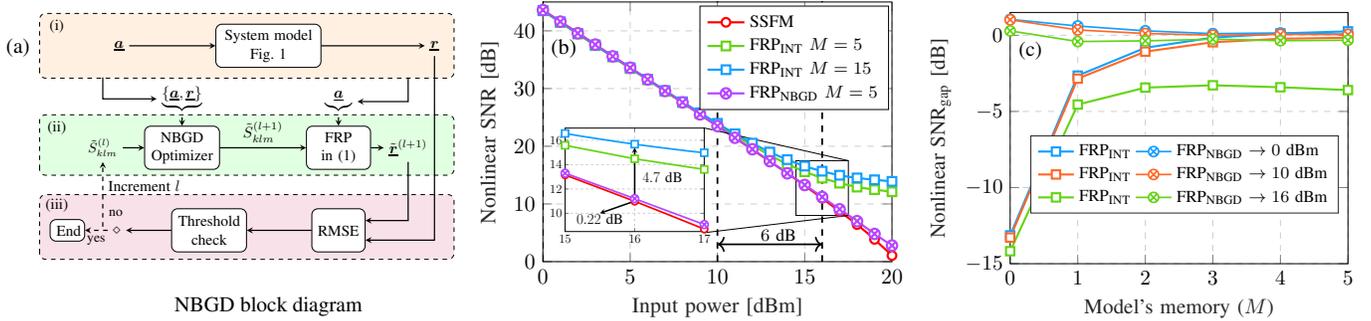}}};

\end{tikzpicture}%
}
 \hfill
 \resizebox{0.32\textwidth}{!}{\definecolor{mycolor1}{rgb}{0.07451,0.62353,1.00000}%
\definecolor{mycolor2}{rgb}{1.00000,0.41176,0.16078}%
\definecolor{mycolor3}{rgb}{0.39216,0.83137,0.07451}%
\begin{tikzpicture}

\begin{axis}[%
every axis/.append style={font=\normalsize},
width=0.4\columnwidth,
height=0.32\columnwidth,
xmin=0,
xmax=5,
xticklabel style={font=\normalsize},
xlabel style={font=\color{white!15!black}},
xlabel={Model's memory ($M$)},
ymin=-15,
ymax=1.5,
ylabel style={font=\color{white!15!black}},
yticklabel style={font=\normalsize},
ylabel={$\text{Nonlinear SNR}_{\text{gap}}\text{ [dB]}$},
axis background/.style={fill=white},
grid style={dashed,lightgray!75},
xmajorgrids,
ymajorgrids,
legend style={at={(0.06,0.2)}, anchor=south west, legend cell align=left, align=left, draw=white!15!black,font=\footnotesize,legend columns=2}
]
\addplot [color=mycolor1, line width=1pt, mark size=2pt, mark=square*, mark options={solid, mycolor1,fill=white}]
  table[row sep=crcr]{%
0	-13.1395378857366\\
1	-2.64781156600484\\
2	-0.821770470918111\\
3	-0.150963161252591\\
4	0.121435846065076\\
5	0.259730149556603\\
};
\addlegendentry{$\text{FRP}_{\text{INT}}$}

\addplot [color=mycolor1, line width=0.8pt,mark=otimes*,mark options={fill=white,scale=1.2}]
  table[row sep=crcr]{%
0	1.0446849186687\\
1	0.611077126922076\\
2	0.292775745662965\\
3	0.10859734611909\\
4	0.150730982466769\\
5	0.0677567706344888\\
};
\addlegendentry{$\text{FRP}_{\text{NBGD}}$ $\rightarrow 0$ dBm}

\addplot [color=mycolor2, line width=1pt, mark size=2pt, mark=square*, mark options={solid, mycolor2,fill=white}]
  table[row sep=crcr]{%
0	-13.2835005136202\\
1	-2.84794947298458\\
2	-1.07173657285436\\
3	-0.456311078099493\\
4	-0.22745950398021\\
5	-0.158421561605902\\
};
\addlegendentry{$\text{FRP}_{\text{INT}}$}

\addplot [color=mycolor2, line width=0.8pt,mark=otimes*,mark options={fill=white,scale=1.2}]
  table[row sep=crcr]{%
0	1.01967279524467\\
1	0.352978096178035\\
2	0.117219798105641\\
3	0.01051698191376\\
4	0.0590971322139495\\
5	0.0512625536079021\\
};
\addlegendentry{$\text{FRP}_{\text{NBGD}}$ $\rightarrow 10$ dBm}

\addplot [color=mycolor3, line width=1pt, mark size=2pt, mark=square*, mark options={solid, mycolor3,fill=white}]
  table[row sep=crcr]{%
0	-14.179178069806\\
1	-4.54813361042918\\
2	-3.43515025985068\\
3	-3.28155635027785\\
4	-3.41126725112372\\
5	-3.59668862445447\\
};
\addlegendentry{$\text{FRP}_{\text{INT}}$}

\addplot [color=mycolor3, line width=0.8pt, mark=otimes*,mark options={fill=white,scale=1.2}]
  table[row sep=crcr]{%
0	0.29181553770902\\
1	-0.409511897363771\\
2	-0.365667582584484\\
3	-0.255982068972507\\
4	-0.352892550525834\\
5	-0.323709825337781\\
};
\addlegendentry{$\text{FRP}_{\text{NBGD}}$ $\rightarrow 16$ dBm}

\node[font=\normalsize] at (0.3,-1) {(c)};

\end{axis}
\end{tikzpicture}%
}\vspace{-.3cm}
 \caption{{\small (a) NBGD block diagram (b) SNR vs launch power, (c) SNR gap with respect to SSFM vs FRP model's memory.}} \label{fig:summary_assessment}
\end{figure*}

As an alternative to computing the FRP coefficients via \eqref{eq:SMP_kernel}, we propose a data-driven method to optimise them using a normalized batch gradient descent (NBGD) algorithm. NBGD is a popular enhancement to the standard batch gradient descent that deals with small gradient magnitudes near stationary points \cite{Watt}. NBGD is flexible to be used with transmission data obtained via either lab experiments or numerical simulations. 
The NBGD algorithm operates minimizing the root mean squared error (RMSE) between a $B$-length batch of true-transmission outputs $\boldsymbol{r}_1,\cdots, \boldsymbol{r}_B$, here split-step Fourier method (SSFM) results, and their corresponding approximation $\tilde{\boldsymbol{r}}_1,\cdots, \tilde{\boldsymbol{r}}_B$ given by \eqref{eq:single_ch_dual_pol_DRP}. Fig.~\ref{fig:summary_assessment} (a) depicts a block diagram for the NBGD, which consists of three stages:  (i) data collection, (ii) gradient step and estimation computation, and (iii) objective function assessment. 

The NBGD performance is studied in terms of the signal to nonlinear-noise ratio (nonlinear SNR). Fig. \ref{fig:summary_assessment} (b) and (c) display the SNR results for a noiseless, $L=120$~km standard single-mode fiber transmission, at $60$~Gbaud, and using DP-$16$-Quadrature Amplitude Modulation(QAM). SSFM results are considered as the benchmark.
Fig.~\ref{fig:summary_assessment}~(b) displays the SNR as a function of the input power, and Fig.~\ref{fig:summary_assessment}~(c) the SNR gap with respect to SSFM as a function of the model's memory size $M$. FRP$_{\text{INT}}$ considers the coefficients as in \eqref{eq:SMP_kernel}, while FRP$_{\text{NBGD}}$ uses the NBGD-estimated coefficients. For the low/intermediate power regime (up to $10$~dBm), Fig.~\ref{fig:summary_assessment}~(b) shows that FRP matches the SSFM results with an accuracy below $0.2$~dB, regardless of how the coefficients were calculated, confirming the FRP good fit for the linear/pseudolinear regime. Nonetheless, Fig.~\ref{fig:summary_assessment}~(c) shows, for the two representative powers of $0$~dBm and $10$~dBm , that the model based on NBGD exhibits early convergence in memory compared to the INT counterpart. For high powers, Fig. \ref{fig:summary_assessment}~(b) shows that FRP$_{\text{INT}}$ diverges from SSFM, a trend that does not improve as the model memory increases, which shows the model lacks accuracy in that regimen. On the other hand, the NBGD optimized coefficients lead to an effective FRP model capable to close the SNR gap, allowing the FRP model to match SSFM results $6$~dB above the power at which the integral-based model starts to fail ($10$~dBm). The inset of Fig.~\ref{fig:summary_assessment} (b) shows that, at $16$~dBm, the gap reduces to approximately $0.22$~dB. As shown in Fig.~\ref{fig:summary_assessment} (c), this extension in the range of validity is achieved with a smaller model's memory. At $M=1$ the effective FRP model shows good accuracy, i.e. gap values close to 0, in all power regimes. This memory reduction leads to a simpler, yet still accurate, FRP model.

\vspace{-0.1cm}
\section{Conclusions}
A data-driven method for the optimization of perturbation coefficients was introduced. This optimization improves the accuracy and extends the validity of first-order regular perturbation models. Future works include extensions to multispan and multi-channel transmission as well as receiver design based on the improved coefficients.
\vspace{-0.2cm}
\bibliographystyle{IEEEtran}
\bibliography{bibliography}

\end{document}